\documentclass[reprint]{revtex4-1}
\usepackage{amsmath}
\usepackage{amssymb}
\usepackage{graphicx}
\usepackage{bm}

\begin{document}
	\title{Turbulence in three-dimensional simulations of magnetopause reconnection}
	
	\author{L.~Price}
	\affiliation{IREAP, University of Maryland, College Park MD 20742-3511, USA}
	\author{M.~Swisdak}
	\affiliation{IREAP, University of Maryland, College Park MD 20742-3511, USA}
	\author{J.~F.~Drake}
	\affiliation{Department of Physics, the Institute for Physical Science
		and Technology and the Joint Space Science Institute,
		University of Maryland, College Park MD 20742-3511, USA}
	\author{J.~L.~Burch}
	\affiliation{Southwest Research Institute, San Antonio, TX 78228, USA}
	\author{P.~A~.Cassak}
	\affiliation{Department of Physics and Astronomy, West Virginia University,
		Morgantown, WV 26506, USA}
	\author{R.~E.~Ergun}
	\affiliation{Department of Astrophysical and Planetary Sciences, University of
		Colorado Boulder, CO 80303, USA}

\begin{abstract}

We present detailed analysis of the turbulence observed in
three-dimensional particle-in-cell simulations of magnetic
reconnection at the magnetopause. The parameters are representative of
an electron diffusion region encounter of the Magnetospheric
Multiscale (MMS) mission.  The turbulence is found to develop around
both the magnetic X line and separatrices, is electromagnetic in
nature, is characterized by a wave vector $k$ given by
$k\rho_e\sim(m_eT_e/m_iT_i)^{0.25}$ with $\rho_e$ the electron Larmor
radius, and appears to have the ion pressure gradient as its source of
free energy. Taken together, these results suggest the instability is
a variant of the lower hybrid drift instability.  The turbulence
produces electric field fluctuations in the out-of-plane direction
(the direction of the reconnection electric field) with an amplitude
of around $\pm 10$~mV/m, which is much greater than the reconnection
electric field of around $0.1$~mV/m. Such large values of the
out-of-plane electric field have been identified in the MMS data. The
turbulence in the simulations controls the scale lengths of the
density profile and current layers in asymmetric reconnection, driving
them closer to $\sqrt{\rho_e\rho_i}$ than the $\rho_e$ or $d_e$
scalings seen in 2-D reconnection simulations, and produces significant
anomalous resistivity and viscosity in the electron diffusion region.

\end{abstract}
\maketitle
%\begin{article}

\section{Introduction}
During magnetic reconnection topological changes in the magnetic field
trigger the transfer of magnetic energy to the surrounding plasma,
where it appears as flows, thermal energy, and nonthermal particles.
The change of topology occurs at magnetic X lines, which are embedded
within electron diffusion regions.  The recently launched
Magnetospheric Multiscale (MMS) mission is designed to make
high-resolution spatial and temporal measurements within electron
diffusion regions and explore the small-scale activity, including
turbulence, found there \citep{burch16a}.

The initial phase of the MMS mission focused on the magnetopause, the
location where the plasmas of the magnetosheath and magnetosphere
reconnect.  Such reconnection is typically asymmetric
\citep{cassak07a} and includes significant differences between the
magnetic fields, densities, and ion and electron temperatures.  The
strong gradients across the magnetopause associated with these asymmetries are susceptible to
the generation of drift waves and their associated instabilities.  Of
particular interest for reconnection, which produces ambient gradients
with scale lengths at or below the ion Larmor radius $\rho_i$ or ion
inertial scale $d_i$, is the lower hybrid drift instability
(LHDI). The theory of this instability has been widely explored in
previous work \citep{davidson75a,daughton03a,huba82a,
  roytershteyn12a, winske81a, yoon08a}.

The fundamental energy sources for LHDI are magnetic field and plasma
pressure inhomogeneities that drive the relative drifts of electrons
and ions.  In the case of the magnetopause the relative drift of the
electrons and protons arises dominantly from the ${E}\times {B}$ drift of electrons: the ion pressure across the magnetopause is
to lowest order balanced by a Hall electric field
\begin{equation}
  E_x\sim\frac{1}{ne}\frac{\partial P_i}{\partial x}\sim\frac{P_i}{neL_i}
\end{equation}
with $P_i$ the ion pressure and $L_{i}$ the ion pressure scale length
(we use GSM coordinates with $x$ pointing toward the sun, $y$ pointing
in the azimuthal direction and $z$ pointing to the north). The
consequence is that to lowest order the net ion drift in the $y$
direction is zero because the ${E}\times {B}$ and diamagnetic
drifts cancel.  The Hall electric field drives a current of electrons
\begin{equation}
  v_{de}\sim\frac{cE_x}{B_z}\sim\frac{cT_i}{eB_zL_i}\sim v_{*i}
\end{equation}
in the $y$ direction that is equal in magnitude to the ion diamagnetic
velocity $v_{*i}=v_i\rho_i/L_i$. This strong drift is reflected in the
crescent-shaped electron velocity distributions documented in MMS
observations \citep{burch16a}. Because $T_i\gg T_e$ at the
magnetopause, electron diamagnetic drifts are small compared with this
${E}\times {B}$ drift. Thus, it is fundamentally the ion
pressure gradient that is the driver of the relative drift of ions and
electrons and the driver of drift-type instabilities.  (This statement
can be cast in a different form by noting that the ion-pressure-driven
drifts and associated currents support the reversal in the
direction of the magnetic field across the magnetopause. Since the
system is inductive, the integral of the current across the reversal
is an invariant and the magnetic free energy, which must be related to
the pressure, can be considered the effective energy source.)

Regardless of the physical description, the basic characteristics of
the LHDI in the low-$\beta$ ``local approximation'' (for which the
profiles of pressure and current are neglected) are electrostatic
oscillations, ${k}\boldsymbol{\cdot}{B}=0$, a most
unstable mode satisfying $k\rho_e \sim 1$, and $\omega \sim kv_{*i}
\lesssim \Omega_{\text{lh}}$. Here $\omega$ is the mode frequency, $k$ is
the wave number, $\rho_e$ is the electron Larmor radius and
$\Omega_{\text{lh}}=\sqrt{\omega_{ce}\omega_{ci}}$ is the lower hybrid
frequency.

However, these properties are modified when the LHDI is excited in a
narrow current sheet (one with a width of order the ion gyroradius or
smaller).  Theory and simulations \citep{winske81a,daughton03a}
suggest that the ``local'' mode described above quickly saturates and
another longer-wavelength instability subsequently develops.  The new
LHDI mode is electromagnetic and has a wave number $k\rho_e \sim
(m_e T_e/m_i T_i)^{0.25}$.  In addition, while the shorter wavelength
electrostatic fluctuations tend to be confined to the edges of the
current sheet (being stabilized at high $\beta=8\pi P/B^2$), the
longer wavelength electromagnetic mode penetrates to the sheet's
center.  Moreover, the electromagnetic mode need not strictly satisfy
${k}\boldsymbol{\cdot}{B}=0$.  In light of this longer wavelength mode, LHDI is expected 
to satisfy somewhat more relaxed conditions: $(m_e T_e/m_i T_i)^{0.25}\leq k\rho_e \leq 1$
and $\omega_{\text{ci}}\leq \omega \leq
\Omega_{\text{lh}}$.

In a previous paper \citep{price16a} we performed a three-dimensional
simulation of reconnection with initial conditions representative of
an MMS observation of an electron diffusion region \citep{burch16a}.
As part of that work we observed turbulence developing around both the
X line and the separatrices. We suggested that the turbulence was due
to LHDI.  These conclusions were consistent with earlier magnetopause
observations \citep{bale02a,vaivads04a} and with the more recent MMS observations
of fluctuations \citep{graham17a}.  Others have noted, however, that the turbulence measured by
MMS did not satisfy the criteria for the ``local'' LHDI outlined above
\citep{ergun17a}.  In this work, we perform a more detailed analysis
of the turbulence produced in reconnection simulations and conclude
that it, in fact, shares many characteristics with the longer
wavelength electromagnetic version of the LHDI. These conclusions are
consistent with \citet{le17a}. In addition, we identify
characteristics of the turbulence in our simulations that are
consistent with MMS observations.

A second important issue is whether the turbulence driven by the LHDI
is strong enough to control both the characteristic scale lengths of
the density and current across the electron diffusion region and the
effective Ohm's law \citep{che11a} that controls large-scale
reconnection. In observations of reconnection in the laboratory
\citep{ren08a} and the magnetosheath \citep{phan07a} electron current
layers were broader than the electron scales expected from the results
of 2-D reconnection simulations \citep{drake08a}. Yet previous 3-D
simulations of asymmetric reconnection relevant to the magnetopause
\citep{pritchett13a,pritchett12a,roytershteyn12a}, while exhibiting
turbulence consistent with the LHDI, suggested that the turbulence was
not strong enough to significantly impact the effective Ohm's law in
the electron diffusion region.  However, in the MMS event of 16 October
2015, the density jumped across the magnetopause by a factor of $17$,
which was substantially larger than considered in these previous
simulations.  \citet{price16a}, in simulations of this
large-density-contrast event, found that the turbulence-induced drag and
viscosity were large enough to impact the effective Ohm's
law. However, others suggested that the turbulence only transiently
remained strong enough to influence the average Ohm's law and that at
late time the effective anomalous resistivity and viscosity were
unimportant \citep{le17a}. None of the simulations carried out to date
have established the characteristic scale lengths of the magnetopause
current layer and density profile.

Thus, in the present manuscript we address some basic questions. Is
the turbulence that develops in simulations of the MMS magnetopause
observations consistent with the long-wavelength LHDI? Is the
turbulence strong enough to impact the effective Ohm's law during
magnetopause reconnection? What is the scaling of the characteristic
current layer width and density scale length across the magnetopause?
Section \ref{sims} presents the details of the simulations, section
\ref{analysis} presents our analysis of the turbulence, and section
\ref{discussion} discusses the results and our conclusions.

\section{Simulations}\label{sims}
We use the particle-in-cell code {\tt p3d} \citep{zeiler02a} to
perform the simulations. Lengths are normalized to the ion inertial
length $d_i=c/\omega_{pi}$, where $\omega_{pi} = \sqrt{4\pi n_0
  e^2/m_i}$ is the ion plasma frequency, and times are normalized to
the ion cyclotron time $\omega_{ci0}^{-1} = m_ic/eB_0$.  A nominal
magnetic field strength $B_0$ and density $n_0$ define the Alfv\'en
speed $v_{A0}=\sqrt{B_0^2/4\pi m_in_0}$ which serves as the velocity
normalization.  Electric fields and temperatures are normalized to
$v_{A0}B_0/c$ and $m_iv_{A0}^2$, respectively.

Two simulations presented here were first discussed in
\citet{price16a}. Their initial conditions mimic the observations by
MMS of a magnetopause diffusion region encounter on 16 October 2015
that is described in \citet{burch16a}.  We use the right-handed $LMN$
coordinate system, in which $L$ is in the direction of the
reconnecting magnetic field (roughly north-south), $N$ parallels the
inflow direction (roughly radial), and $M$ (roughly azimuthal) is
perpendicular to $L$ and $N$ in the out-of-plane direction.  The
particle density $n$, reconnecting magnetic field component $B_L$, and
ion temperature $T_i$ vary as functions of $N$ with hyperbolic tangent
profiles of width 1.  The asymptotic values of $n$, $B_L$, and $T_i$
are 1.0, 1.0, and 1.37 in the magnetosheath and 0.06, 1.70, and 7.72
in the magnetosphere.  The small guide field, $B_M=0.099$, is
initially uniform.  The profile of the electron temperature $T_e$ is
determined by pressure balance, with the asymptotic values fixed to
0.12 in the magnetosheath and 1.28 in the magnetosphere.  While in
pressure balance this choice of initial conditions is not a Vlasov
(kinetic) equilibrium.  However, any evolution due to this lack of
equilibrium is quickly overwhelmed by the development of reconnection
and turbulence.

We perform two three-dimensional simulations of this system, with
computational domains of dimensions $(L_L,L_M,L_N) =
(40.96,10.24,20.48)$ and $(20.48,5.12,10.24)$, respectively. These
simulations have the same asymptotic plasma parameters and only differ
in computational parameters, namely the ion-to-electron mass ratio,
the grid resolution, and the speed of light.  The mass ratios are
chosen to be 100 and 400, respectively, which eases the computational
expense associated with using the true mass ratio yet is sufficient to
separate the ion $d_i$ and electron $d_e$ scales ($d_e=0.1d_i$ and
$0.05d_i$, respectively).  Note that although the computational
domains differ in size when measured in $d_i$, they are the same size
when measured in electron scales ($d_e$ or $\rho_e$).  We also
performed a companion two-dimensional simulation with identical
parameters and dimensions $(L_L,L_N)=(40.96,20.48)$.

The spatial grids have resolutions of $\Delta=0.02$ and $\Delta=0.01$,
respectively, which resolve the system's smallest physical scale, the
Debye length in the magnetosheath, $\approx0.03$.  We use 50 particles
per cell per species when $n=1.0$ and, as this implies $\approx3$
particles per cell in the low-density magnetosphere, our analysis
employs, when necessary, averages over multiple cells to mitigate the
resulting noise.  The speed of light is chosen to be $c=15$ and $30$
in the respective simulations, and our boundary conditions are
periodic in all directions.  While periodic boundary conditions
present some limitations, the perturbations observed in our
simulations propagate only a short distance during the length of the
simulation, which suggests that the periodicity in the $M$ direction
has no adverse effect.  A small perturbation is added to initialize
reconnection.  Companion two-dimensional simulations show that
reducing the size of this perturbation by a factor of 2 has no
significant effect other than delaying the onset of reconnection.
Unless otherwise stated, the subsequent figures and discussion focus
on the larger three-dimensional simulation with $m_i/m_e = 100$.

\section{Analysis}\label{analysis}

\begin{figure}
	\centering
	\includegraphics[width=0.9\columnwidth]{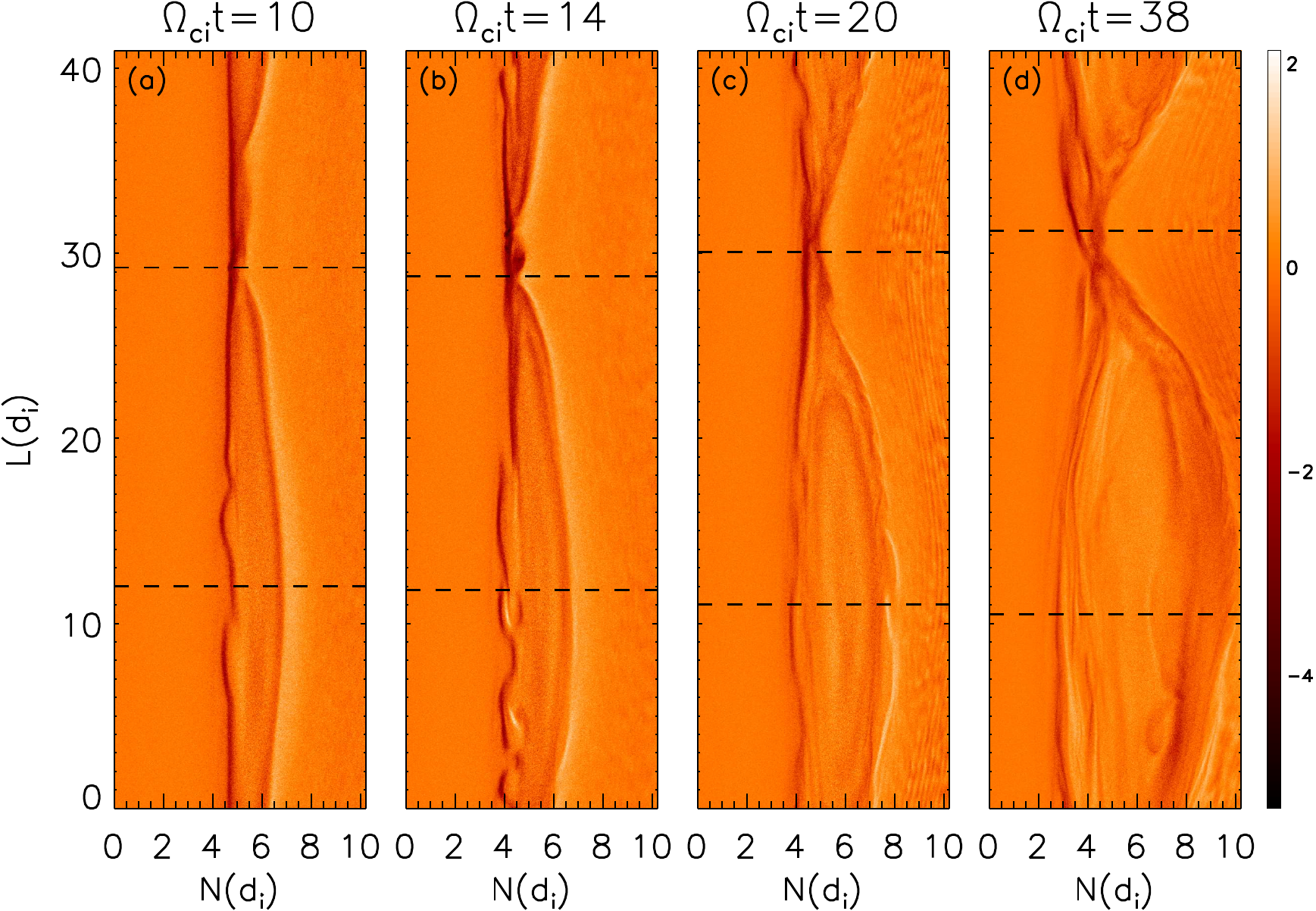}
	\caption{\label{jez} Snapshots of $J_{eM}$, the dawn-dusk
          electron current density, in one $L-N$ plane.  (a--d)
          Taken at $t=10$, $14$, $20$, and $38$, respectively.
          These times highlight the onset of the instability, a time
          of maximum growth, a decrease in power, and the end of the simulation.  The colors in each
          panel are identically normalized, with the color bar at the
          right showing the range.  The dashed lines in each panel
          indicate the locations of cuts through the X line and island
          presented in Figure \ref{ezbx}.}
\end{figure}

In two-dimensional simulations, where variations in the out-of-plane
($M$) direction are suppressed, reconnection in this system remains
laminar \citep{price16a}.  In contrast, the additional freedom present in
three-dimensional simulations allows modes to develop with finite
$k_M$. Figure \ref{jez} displays images of $J_{eM}$, the dawn-dusk
electron current density, in a single $L-N$ plane at four
representative times.  The reason for choosing these times will
be discussed further below, but they roughly correspond to the onset
of the instability, a time of maximum growth, a decrease in power, and the end of the simulation.  The
magnetosphere (strong field, low density, and high temperature) is to the
left and the magnetosheath (weak field, high density, and low temperature)
to the right.  The results exhibit the typical features of asymmetric
reconnection, including the bulge of the magnetic islands into the
low-field-strength magnetosheath and the separation between the
$x$ point and the stagnation point of the fluid flow
\citep{cassak07a,price16a}. As can be seen in Figure~\ref{jez}(a), turbulence
first develops along the magnetospheric separatrix before developing
at the X line (Figure~\ref{jez}b) and the magnetosheath separatrix (Figures~\ref{jez}c
and \ref{jez}d).  Images from other $L-N$ planes exhibit similar features.

\begin{figure}
	\centering
	\includegraphics[width=0.9\columnwidth]{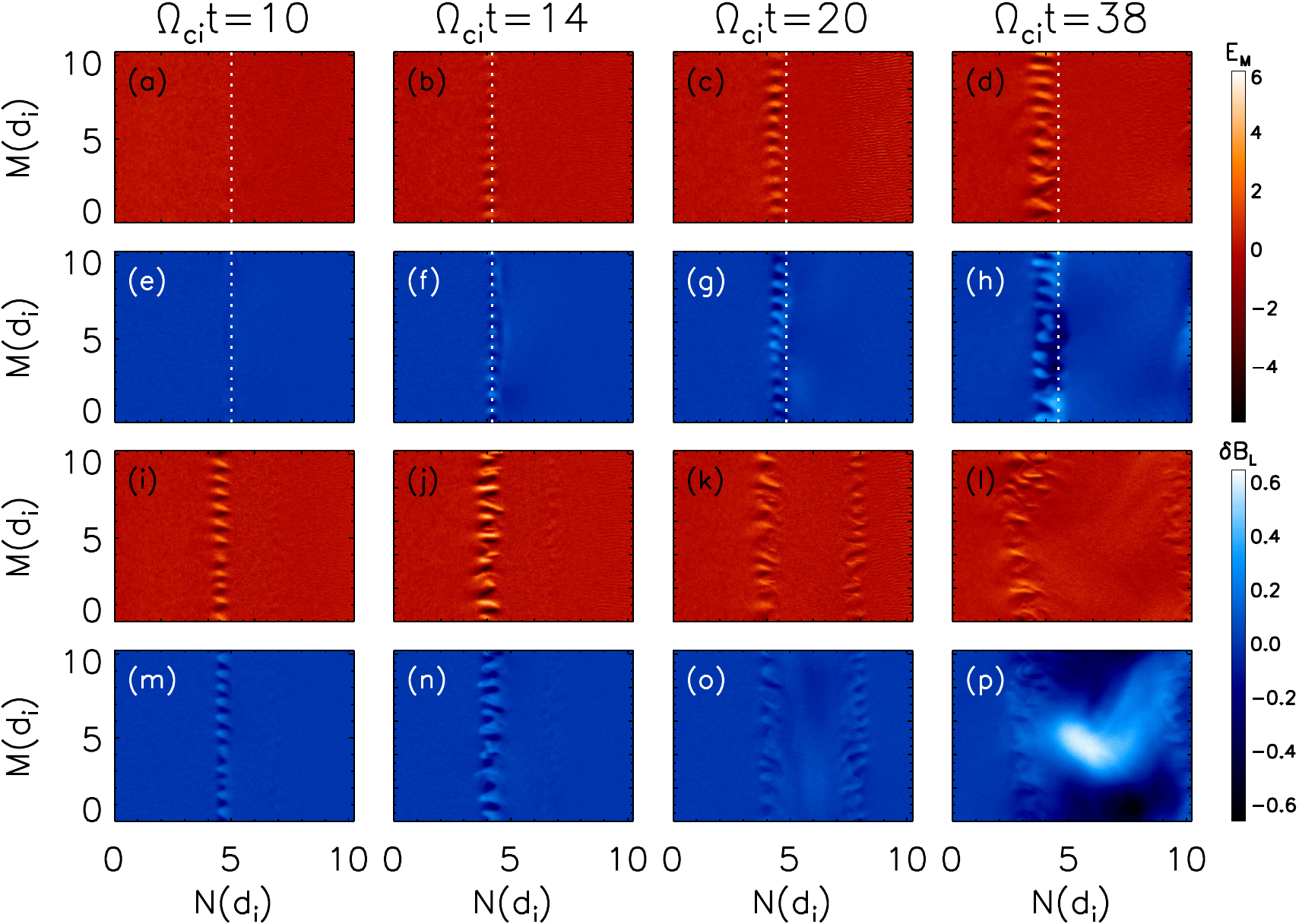}
	\caption{\label{ezbx} Snapshots of (a--d, i--l) $E_{M}$, the electric field
          in the direction of the reconnection-associated current, and
          (e--h, m--p) $\delta B_{L}$, the fluctuations in the reconnecting magnetic
          field, in the $M-N$ plane at the same times as in Figure
          \ref{jez}.  The cuts were taken at the positions shown by
          the dashed lines in Figure \ref{jez}.  Figures~\ref{ezbx}a--\ref{ezbx}h are
          taken at the $L$ location of the x-line, while Figures~\ref{ezbx}i--\ref{ezbx}p
          are taken through the middle of the island.  The red color
          bar corresponds to $E_M$, while the blue color bar
          corresponds to $\delta B_L$.  The dotted lines in Figures~\ref{ezbx}a--\ref{ezbx}h
          correspond to the $N$ location of the X line.}
\end{figure}

The instability driving the turbulence is electromagnetic in nature,
as can be seen in Figure \ref{ezbx}.  Figures~\ref{ezbx}a--\ref{ezbx}h show $E_M$ and
$\delta B_L$ in the $M-N$ plane that cuts through the X line, while
Figures~\ref{ezbx}i--\ref{ezbx}p show the same quantities along a cut through the island.
Here, $\delta B_L$ is the fluctuating component of $B_L$, that is,
$\delta B_L=B_L-\langle B_L \rangle$, where $\langle B_L \rangle$ is
$B_L$ averaged over the $M$ direction.  This is the dominant magnetic
field perturbation---convection of the large gradient of $B_L$ in the
initial state due to the perturbed $v_{eN}$ leads to large
fluctuations.  Fluctuations of $B_M$ and $B_N$ are also present but at
a reduced amplitude \citep{price16a}.

The turbulence first appears in both $E_M$ and $\delta B_L$ at $t=10$
along the magnetospheric separatrix in Figures~\ref{ezbx}i and \ref{ezbx}m.  Turbulence
develops at the X line (Figures~\ref{ezbx}b and \ref{ezbx}f) and along the
magnetosheath separatrix (Figures~\ref{ezbx}j and \ref{ezbx}n) by $t=14$, though the
latter is clearer by $t=20$ (Figures~\ref{ezbx}k and \ref{ezbx}o).  It is interesting
to note that even at relatively early times, the location of the
turbulence begins to shift away from the X line, denoted by the white
dotted lines in Figures~\ref{ezbx}a--\ref{ezbx}h, toward the magnetosphere.  We also
observe evidence of a possible kink mode late in the simulation in
Figure~\ref{ezbx}p. This mode produces a global perturbation to the current
sheet, but at longer wavelength than the fluctuations seen in the
other panels.
 
\begin{figure}
	\centering
	\includegraphics[width=0.9\columnwidth]{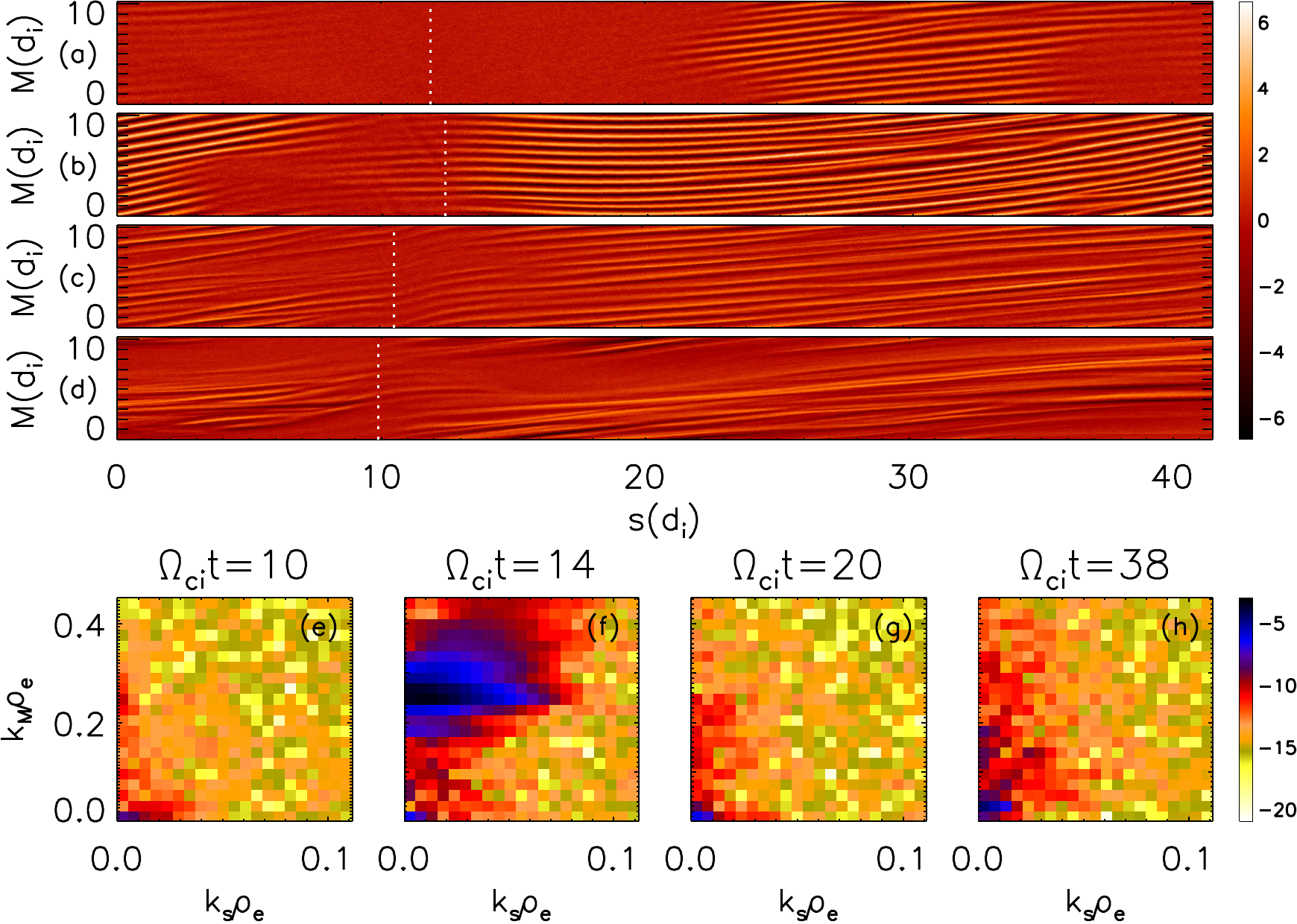}
	\caption{\label{fftez} (a--d) $E_M(s,M)$ and (e--h) Fourier
          transforms at the same times as in Figure
          \ref{jez}, where $s$ is the distance along the average (over
          $M$) magnetic field. The data are from a surface that lies to
          the left (magnetospheric side) of the magnetospheric
          separatrix. Figure~\ref{fftez}a shows $E_M(s,M)$ at $t=10$; the
          $k_s-k_M$ power spectrum $\log(|\tilde{E}_M(k_s,k_M)|^2)$ at
          the same time is shown in Figure~\ref{fftez}e.  Figures~\ref{fftez}f, \ref{fftez}g, and \ref{fftez}h are similarly paired with Figures~\ref{fftez}b, \ref{fftez}c, and \ref{fftez}d and show the simulation data at $t=14$, $20$, and $38$,
          respectively. The white dotted lines in Figures~\ref{fftez}a--\ref{fftez}d
          correspond to the location of the X line.  Figures~\ref{fftez}a--\ref{fftez}d are
          normalized to the same value, as seen in their accompanying
          color bar.  Figures~\ref{fftez}e--\ref{fftez}h also have a common normalization.}
\end{figure}

The wavelength of the drift instability can be directly measured in
several of the panels.  In Figure~\ref{ezbx}b, for example, there
are 11 wavelengths present in the $M$ direction (length $10.24 d_i$). The choice of
temperatures to use in the conversion from $d_i$ to $\rho_e$ is somewhat arbitrary due to
the strong gradients in the system and the fact that the instability is a global mode across the magnetopause and along the local magnetic field (see Fig.~\ref{fftez}).  In this paper, since most of the plasma at
the X line comes from there, we choose the asymptotic magnetosheath
values.  Other choices can change $\rho_e$ by up to a factor of 2.
Thus, at $t=14$ in our mass ratio 100 simulation $1 d_i\approx 27\rho_e$, which gives
$k_M\rho_e\approx$ 0.25.  As will be discussed later, this is
consistent with the expectation for long wavelength LHDI.

While LHDI is the most likely candidate to explain the
turbulence seen in our simulations, the modified-two-stream instability
(MTSI) can also exist in finite $\beta$ systems if the relative
cross-field drifts of the electrons and ions are comparable to or exceed
the local Alfv\'en speed \citep{wu83a}. It has been suggested that
this instability is important in laboratory reconnection experiments
\citep{Ji04a}. This instability has a growth rate that peaks with a
nonzero component of the wave vector along the local magnetic field
$k_\parallel$ in contrast with the LHDI, which has a peak growth rate
for $k_\parallel=0$ \citep{daughton03a}. Thus, to distinguish between
the possible drivers of the turbulence, we examine its Fourier
spectrum perpendicular to and along the local magnetic field in
$k_\perp - k_\parallel$ space, where $k_\perp$ is calculated from the
data along the $M$ direction. Since the local direction of the
magnetic field varies in space, the necessary data must be taken while
following a magnetic field line. Furthermore, since the actual field
lines have chaotic trajectories \citep{price16a}, the analysis is
carried out using the magnetic field components obtained by averaging
over the $M$ direction. The averaged magnetic field on the
magnetospheric side of the reconnection layer follows the separatrix
between the upstream and reconnected plasma, while $M$ points in the
perpendicular direction. Choosing $s$ to represent the distance
measured along the field, we construct $E_M(s,M)$ while traveling
along a field line just outside the separatrix.  The range of $s$ is
chosen in order to travel through the simulation domain in the $L$
direction exactly once. These data are not periodic in $s$ for a given
value of $M$, but the data can be extended arbitrary distances along
$s$ by stacking the data along $s$ if it is shifted a fixed distance
in $M$.

The resultant $E_M(s,M)$ at four times can be seen in Figures~\ref{fftez}a--\ref{fftez}d.  The primarily horizontal stripes correspond to the
same instability shown in Figure \ref{ezbx}.  In Figure~\ref{fftez}a, calculated
at $t=10$, the instability is weak at the location of the X line (the
white dotted line) but strong near the middle of the island (see
Figure \ref{jez}a). By $t=14$ in Figure~\ref{fftez}b the instability
is present at all values of $s$, including at the X line, although it
remains strongest near the middle of the island.  This pattern
persists at later times, $t=20$ and $38$, Figures~\ref{fftez}c and \ref{fftez}d,
respectively, making it appear that the turbulence near the X line is
not strong.  However, note that as seen in Figures~\ref{ezbx}c and
\ref{fftez}d, the turbulence at these times is displaced from the
separatrix. Although not shown here, $E_M(s,M)$ at the X line is much
stronger along a trajectory that is displaced toward the magnetosphere
compared with that shown in Figure \ref{fftez}.

To determine the dominant wavelengths present in $E_M(s,M)$, we
construct two-dimensional spatial Fourier transforms (denoted by the
operator $\mathcal{F}$) of the $s-M$ domain, $\tilde{E}_M(k_s,k_M) =
\mathcal{F}[E_M(s,M)]$.  We plot $\log(|\tilde{E}_M(k_s,k_M)|^2)$ for
the longest wavelength modes in Figures~\ref{fftez}e--\ref{fftez}h.  At $t=10$,
Figure~\ref{fftez}e, which is the linear stage of the instability, the spectrum
is dominated by nearly perpendicular wave vectors (note the difference
in vertical and horizontal axis scales). The peak power when the
instability is strongest, $t=14$, occurs for $k_M \rho_e~\approx0.25$,
consistent with the calculation based on Figure \ref{ezbx}.  By this
time the spectrum has acquired a significant parallel wavevector
($k_s$), although it continues to be dominated by perpendicular modes.
After saturation (Figures~\ref{fftez}g and \ref{fftez}h), however, those parallel modes
diminish in strength.  Since this simulation employs an
ion-to-electron mass ratio of $100$, theory suggests that the longer
wavelength LHDI mode has $k_M
\rho_e\sim(m_eT_e/m_iT_i)^{0.25}\approx0.14$.  As before, we employ the asymptotic magnetosheath
temperatures since LHDI is a global mode.  The expected value is consistent with our measured
value of $k_M\rho_e\approx0.25$.
%(This comparison neglects the fact that $T_i \neq T_e$ which will
%have a modest effect on the theoretical value.)

The nonlocal structure of the MTSI has not been explored in the 
literature. Nevertheless, in local models the instability peaks at
$k_\parallel/k_\perp \sim \sqrt{m_e/m_i}$ \citep{wu83a}. For the
simulation data shown in Fig.~\ref{fftez} in which $m_i/m_e=100$ the
spectrum should exhibit a distinct peak centered on $k_s\sim0.1k_M$ if
it were driven by the MTSI. There is no evidence for a peak at finite
$k_\parallel$ in the data of Fig.~\ref{fftez}.

However, the data of Fig.~\ref{fftez} do reveal that $k_s$ is
finite. We suggest that this is a consequence of the
inhomogeneity of the out-of-plane current with distance along the
separatrices. As discussed in \citet{price16a}, this instability
dominantly drives flows in the $M-N$ plane. The resulting twisting of
flux ropes by the vortical $M-N$ flows is similar to that inferred
from MMS observations by \citet{ergun16a}. The strength of the
vortices varies with distance along the field line ($s$ direction)
because the amplitude of the out-of-plane current $J_{eM}$ depends on
the distance from the x-line. As a consequence, the rate of twist of
the flux tubes varies with distance from the X line, generating
nonzero values of $B_M$ and $B_N$ and a finite $k_s$.

%\begin{figure}
%	\centering
%	\includegraphics[width=0.9\columnwidth]{power_crop}
%	\caption{\label{pow} Panel (a): Power in the instability based
%          on the Fourier transforms shown in Figure \ref{fftez}.
%          Panel (b): Density profiles at the x-line over time.  Since
%          the location of the density gradient changes as the islands
%          expand, the density profiles at later times are shifted in
%          $N$ to ease comparisons.}
%\end{figure}
%
%
%The total power in the instability's fluctuating electric field
%\begin{equation}\label{poweq}
%P=\sum_{k_s,k_M}|\tilde{E}_M(k_s,k_M)|^2
%\end{equation}
%is computed for the duration of the simulation and plotted in Figure
%\ref{pow}(a).  The instability first becomes noticeable around $t=10$,
%climbs rapidly to a peak at $t=14$ and then rapidly decreases in
%intensity as it non-linearly saturates due to the relaxation of the
%driving gradients.  A secondary instability, perhaps the kink mode
%that can be seen in the final panels of Figure \ref{ezbx}, begins to
%grow, albeit more weakly, near the end of the simulation.  This
%general pattern is not specific to the choice of $E_M$ and is found in
%similar calculations performed on any component of the fluctuating
%electric and magnetic fields.

\begin{figure}
	\centering
	\includegraphics[width=0.9\columnwidth]{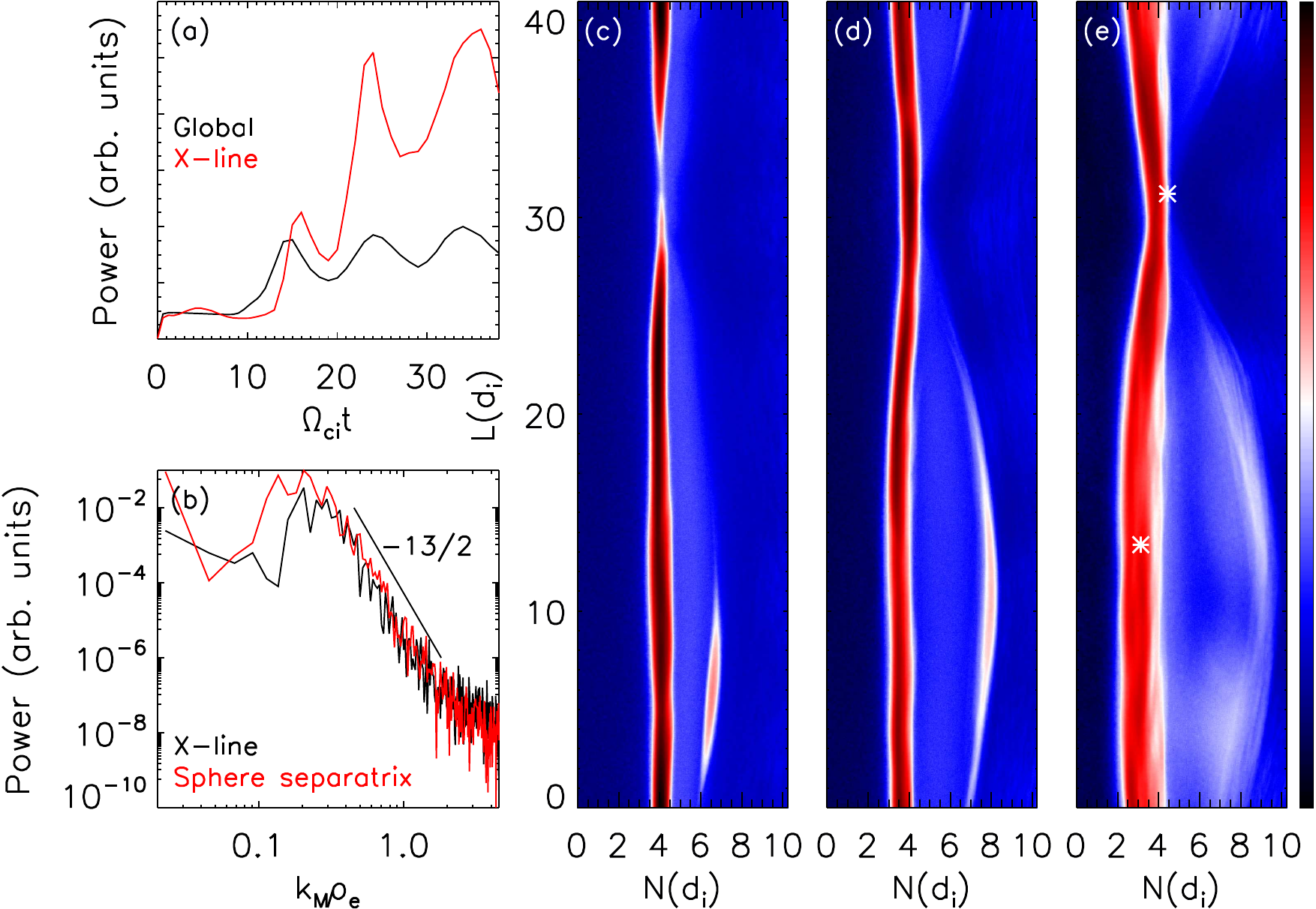}
	\caption{\label{pow} (a) Power in the instability as computed by equation~\eqref{poweq}, based on the fluctuating component of $\delta E_M=E_M-\langle E_M \rangle$ with $\langle\rangle$ denoting an average over the $M$ direction.  The black line corresponds to the global power, while the red line corresponds to the power locally near the X line.  (b) The one-dimensional power spectrum $\lvert\mathcal{F}\left[E_M(M)\right]\rvert^2$ at $t=34$, at the locations denoted by asterisks in Figure\ref{pow}e. The slope of the power law is around $-6.5$. (c--e) $\log \left( \sum_M (\delta E_M)^2 \right)$ at $t=14, 24,$ and $34$, respectively.  Figures~\ref{pow}c--\ref{pow}e have a common normalization.}
\end{figure}
The total power in the instability's fluctuating electric field $\delta E_M=E_M-\langle E_M\rangle$,
\begin{equation}\label{poweq}
P=\sum_{L,M,N} (\delta E_M)^2,
\end{equation}
where $\langle\rangle$ denotes an average over the $M$ direction, is
computed for the duration of the simulation and plotted in Figure
\ref{pow}a.  The power is calculated both globally (the black curve) and for a small region around the X line ($N\in (2,5)$, $L\in (27,36)$, the red curve).  The global power begins to increase at $t=10$,
first peaks at $t=14$, and decreases to a local minimum at $t=20$,
before reaching new peaks at $t=24$ and $34$. The same pattern is observed in the power at the X line, albeit offset slightly in time.  This is consistent with Figures \ref{jez} and \ref{ezbx}, with the instability first appearing along the magnetospheric separatrix before developing at the X line.  The overall increase in the baseline (nonoscillatory) power seen at the X line is due to the spreading of the turbulence over a greater spatial domain and not to an increase in the turbulence's amplitude; the amplitude saturates around $t=14$.  The relative magnitudes of the two curves are not significant.  Instead, what is important is their profile in time.  The periodicity observed
here corresponds to a slow oscillation, or ``breathing,'' of the
current sheet in the $N$ direction and is also observed in
calculations of the reconnection rate (not shown).  This ``breathing''
is a consequence of the absence of a kinetic equilibrium in the
initial state. Figure \ref{pow}b shows the one-dimensional power
spectrum $\lvert\mathcal{F}\left[E_M(M)\right]\rvert^2$ at $t=34$ at
both the X line and the magnetosphere separatrix. The spectrum takes
the form of a power law with a slope of around $-6.5$, which is the same
at both the X line and the separatrix. A power law in the frequency
spectrum of turbulence associated with the LHDI has been seen in the
Polar spacecraft data at the magnetopause \citep{bale02a} although the
spectral index was much harder, $-1$, in comparison to the spectrum
here.  Figures~\ref{pow}c--\ref{pow}e show $\log \left( \sum_M (\delta E_M)^2
\right)$ at the times of the three peaks, representing the strength of
the fluctuations in $E_M$.  As in Figure \ref{ezbx}, the fluctuations
first appear strongest along the magnetosphere separatrix (Figure~\ref{pow}c)
but also appear weakly along the magnetosheath separatrix. At late
time the fluctuations are also evident within the reconnection exhaust
as turbulence around the X line and magnetic separatrices is carried
into the exhaust. A filamentary exhaust has been documented in MMS
observations \citep{phan16a}.

\begin{figure}
	\centering
	\includegraphics[width=0.9\columnwidth]{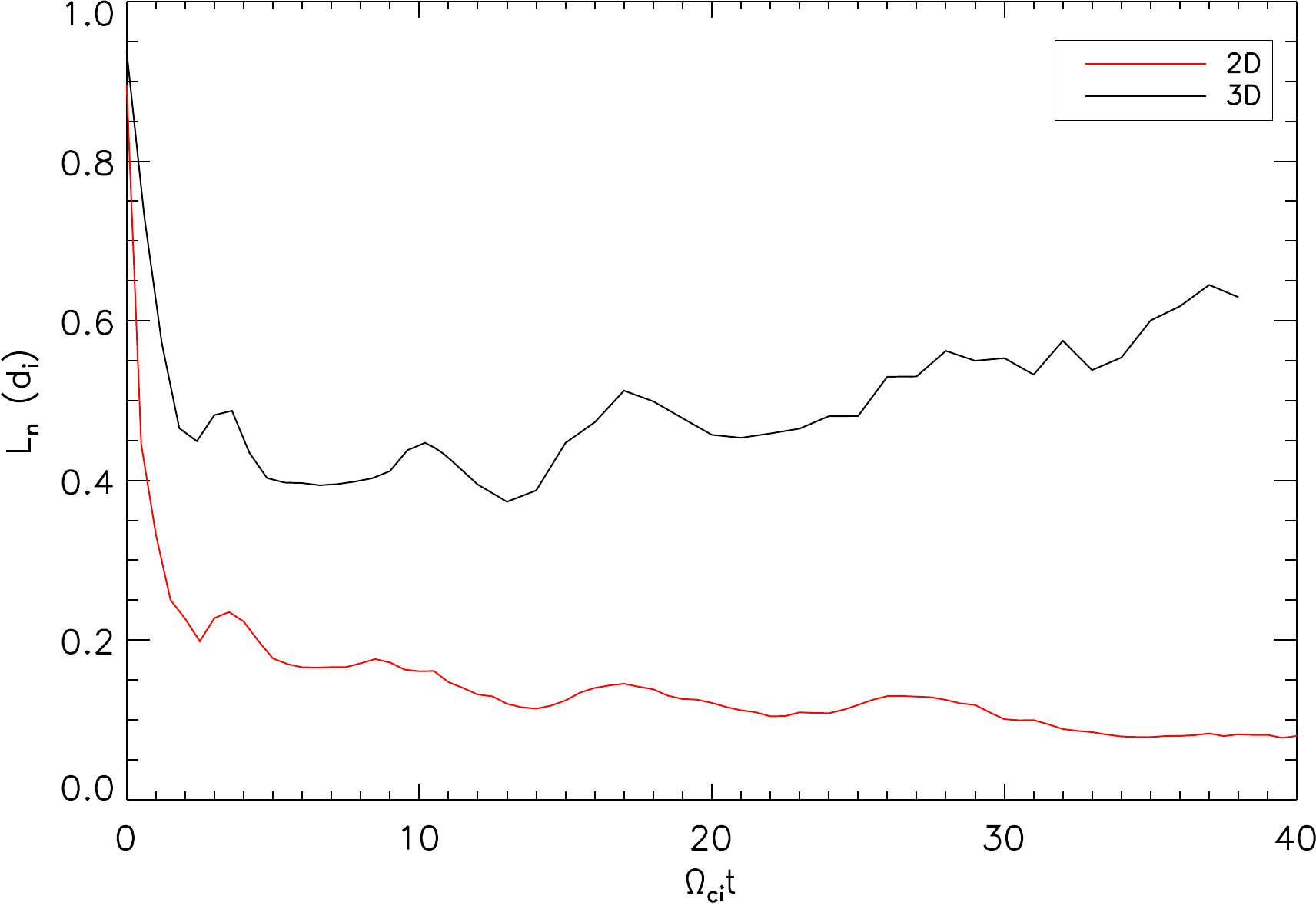}
	\caption{\label{den} The density scale length $L_n$ as a function of time for the mass ratio 100 simulations.}
\end{figure}

As discussed earlier, the energy source for the instability is the
relative electron-ion drift, which is dominantly produced by the ion
pressure gradient. Because of the large drop in the density across the
magnetopause for the initial conditions of the present simulation, the
ion pressure drop is dominated by the change in density. We therefore
explore the linkage between the time evolution of the density profile
and the development of the turbulence to demonstrate the causal
relation between the local gradient and the turbulence. Figure
\ref{den} shows the density scale length at the X line as a function of time for the 3-D and 2-D mass-ratio 100 simulations.

%Because the $N$ location of the current sheet changes as
%reconnection occurs, the density profiles have been shifted to make
%comparisons easier.
For the 3-D simulation, the initial density scale length $L_n$ ($\approx
1 d_i = 10d_e$) decreases as reconnection develops,
reaching its minimum value ($L_n\approx 0.4 d_i= 4 d_e$) at $t=13$, near 
when the instability is strongest.  The density profile then relaxes somewhat
and by the end of the simulation $L_n\approx 0.6 d_i = 6d_e$. Similar behavior is observed for a 3-D simulation with mass ratio 400 (not shown).  This
result should be contrasted with the results of the 2-D simulation with
the same parameters, in which turbulence does not develop
and in which the density gradient steepens in time and comes to a
constant density scale length of around $L_n \approx 0.1 d_1=1.0d_e$. Thus, the turbulence
clearly limits the minimum density scale length and the corresponding
width of the electron current layer.  We note that because of their high cadence, the MMS spacecraft instruments reveal the local density in a cut across the magnetopause rather than the average scale length shown in Fig.~\ref{den}. However, the rate of large-scale reconnection is controlled by the $M$-averaged properties of the system, including the $M$-averaged density. This point will be emphasized below in the discussion of averaged and local Ohm's law for this event.

\begin{figure}
	\centering
	\includegraphics[width=0.9\columnwidth]{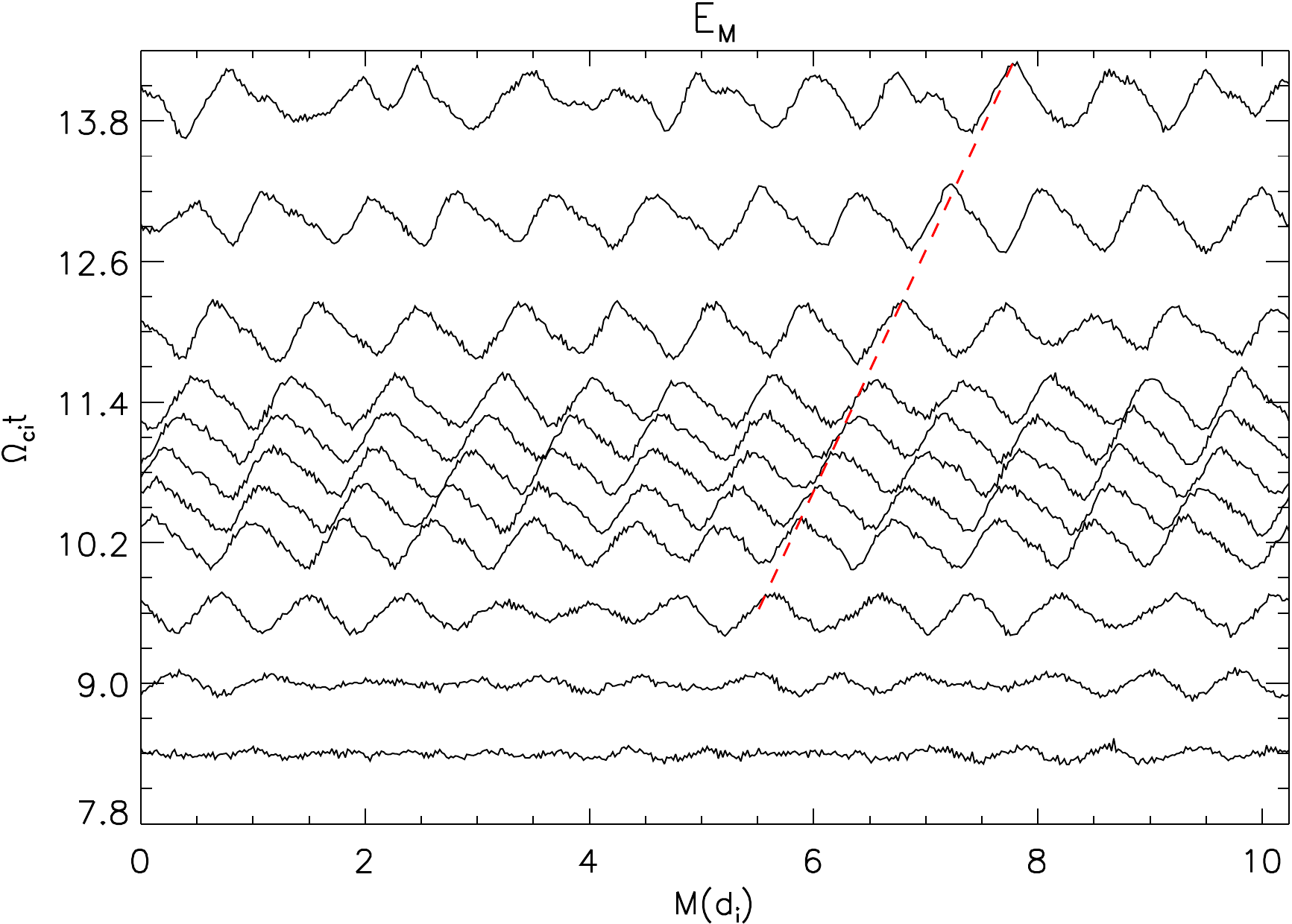}
	\caption{\label{ezstack} Cuts of $E_M$ along the $M$ direction
          through the center of the turbulence at the magnetospheric
          separatrix. The vertical position of each cut is shifted
          based on the time at which it was taken.  The red dashed
          line traces the displacement of one wave peak.}
\end{figure}

Next we calculate the phase speed and frequency of the instability.
Figure \ref{ezstack} shows cuts of $E_M$ along the $M$ direction
through the center of the turbulence at the magnetospheric separatrix
near the middle of the island.  The vertical position of each cut
corresponds to the time at which it was taken.  The turbulence begins
to appear over the background variations at $t\approx 9$ and by $t=10$
has clearly developed linear oscillations.  The topmost trace, at
$t\approx 14$, is taken when the instability is strongest.  The
irregular variations show that it has already reached a nonlinear
stage, and by this time the wave potential is larger than the electron thermal energy.
By tracing the displacement of one wave peak (the red dashed
line), we determine the phase velocity of this wave to be $v_p \approx
\frac{1}{2} v_{A}$ in the direction of the electron diamagnetic drift.  This value is not specific to the wave peak chosen; similar results are obtained by translating the red dashed
line in the $M$ direction to adjacent peaks.  Thus we can compute the
instability frequency in the frame of the simulation $\omega=v_p
k_M\approx 0.25 \Omega_{\text{lh}}$.

This differs significantly from $\Omega_{\text{lh}}$, which is the
textbook frequency of the LHDI. There are two reasons for this.  The
first is that, as discussed in \citet{daughton03a}, electromagnetic
LHDI modes are not fixed at $\Omega_{\text{lh}}$ but can instead have
frequencies anywhere in the range $\omega_{\text{ci}}\leq \omega \leq
\Omega_{\text{lh}}$.  Second, the standard derivation of the frequency
of LHDI is performed in a frame with $E_N=0$, which is not
the case at the magnetopause and is not true for our simulation.  In
the $E_N=0$ frame, the ions have the strongest drift, of the order of
the ion diamagnetic drift velocity (which exceeds the electron
diamagnetic velocity because the ions are hotter than the electrons).
In our system, the ions are close to stationary, so the observed
frequency is naturally lower than the lower hybrid frequency found in
the typical analysis. Further, the mode propagates in the electron
direction, which is consistent with MMS observations of fluctuations
at the magnetopause \citep{graham17a}. In our simulation it is not possible to completely
transform away $E_N$ since this would require $cE_N/B_L$ to be a
constant. It is possible, however, to transform our simulation results
into a frame in which the value of $E_N$ is greatly reduced.  At the
magnetospheric separatrix during the time of linear evolution,
$c({E}\boldsymbol{\times}{B})_M/B^2 \approx cE_N/B_L$
has a peak value of around $ -1.7 v_A$.  In a frame with this
velocity, the phase speed of the wave is $\approx 1.2 v_A$, giving a
frequency of $\omega = 0.6\Omega_{\text{lh}}$, closer to the expected
value.  

\begin{figure}
	\centering
	\includegraphics[width=0.9\columnwidth]{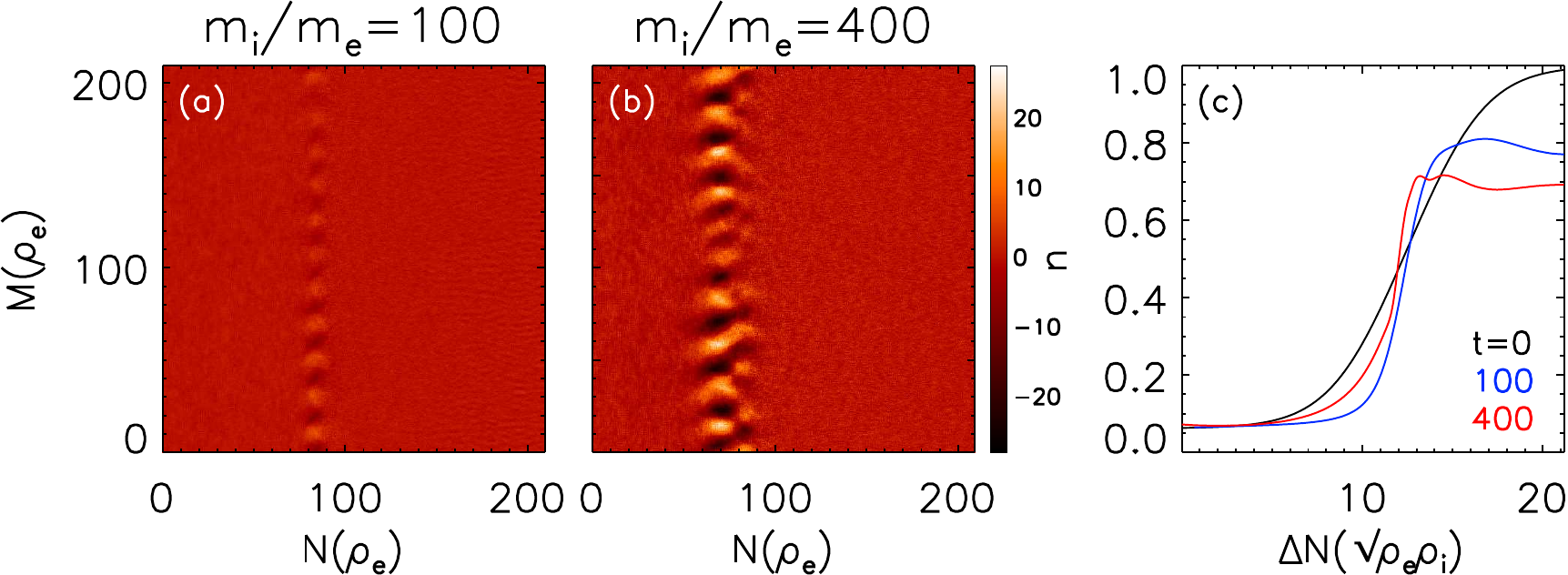}
	\caption{\label{emass} Comparison between the $m_i/m_e=100$ and
          $400$ simulations.  (a--b) Snapshots of $E_M$ in the
          $M-N$ plane through the X line at times of maximum power for
          $m_i/m_e=100$ (Figure~\ref{emass}a) and $400$ (Figure~\ref{emass}b). The
          numerical values of $E_M$ have been converted to units of
          mV/m.  (c) Density profiles at the X line at $t=0$
          and times of minimum density scale length.  The density profiles have been shifted in $N$ to
          facilitate comparison.}
\end{figure}

In \citet{price16a}, we suggested that the qualitative features of a
real mass ratio simulation would not differ significantly from one
with $m_i/m_e=100$.  Although we find that conclusion still holds,
there are important quantitative differences between the simulation
discussed in detail above (mass ratio of 100) and one with mass ratio
400.  Figures~\ref{emass}a and \ref{emass}b show $E_M$ in the $M-N$ plane through
the X line for mass ratio 100 (Figure~\ref{emass}a) and 400 (Figure~\ref{emass}b) at times
of maximum power (as determined using equation \eqref{poweq}).  While
the simulation domains differ in size when measured in $d_i$, they are
equivalent when measured in $\rho_e$.  The instability is stronger in
the mass ratio 400 case and the turbulence has a greater spread in the
$N$ direction.  As before, the wavelength of the instability can be
visually determined.  In Figure \ref{emass}b there are 10
wavelengths in the $M$ direction (length $\approx282 \rho_e$), giving
$k_M\rho_e\approx0.22$.  In agreement with theoretical expectations
there are fewer wavelengths (10 versus 11) and smaller $k_M\rho_e$ for
the more realistic mass ratio.  Furthermore, by constructing
$E_M(s,M)$ and $\log(|\tilde{E}_M(k_s,k_M)|^2)$ (not shown), we find
that the peak of the instability occurs at $k_M\rho_e\approx0.22$.
For an ion-to-electron mass ratio of 400, the longer wavelength LHDI
mode is expected to satisfy
$k_M\rho_e\sim(m_eT_e/m_iT_i)^{0.25}\approx0.09$. Note though that as
discussed below, the ambient density gradient also varies between the
two simulations so the scaling $k_M\rho_e\sim(m_eT_e/m_iT_i)^{0.25}$ is only
approximate.

The scale lengths of the density and current layers at the
magnetopause are topics of scientific interest since they are linked
to the processes that limit the electron current. As noted previously,
our 2-D simulations show that density scale length is of order $1d_e$,
which is the expected value during reconnection without turbulence. The current
layers in the 3-D simulations are limited by the development of
turbulence and never reach electron scales. Because our simulations
are carried out with artificial mass ratios, care must be taken in
interpreting the data. In Figure~\ref{emass}c we display density
profiles at the X line for our
mass ratio $100$ and $400$ simulations.  The initial density profile
is the same for both simulations.  The profiles displayed for each
mass ratio are chosen to correspond to the time when the density
gradient is greatest.  The horizontal length scale is measured in
hybrid units, $\sqrt{\rho_e\rho_i}$. Thus, the minimum scale length of
the density profile (and the current profile) during reconnection at
the magnetopause appears to scale as the hybrid of the electron and
ion Larmor radii rather than either the electron or ion
scale. However, because of the weak dependence of this scaling on the
mass ratio and the limited mass ratios explored in the simulations,
there is some uncertainty in this conclusion. Nevertheless, the current and density scale lengths at the
magnetopause are significantly greater than the expected $d_e$ or
$\rho_e$ scale. Further, the widths are comparable to measurement of
the widths of current layers during symmetric reconnection in the
magnetosheath \citep{phan07a} and in a laboratory reconnection
experiment \citep{ren08a}.  Consistent with the simulation results the
analysis of MMS observations at the magnetopause also suggested that
such turbulence was responsible for electron transport across the
X line from the magnetosheath into the magnetosphere
\citep{graham17a}.

\section{Discussion}\label{discussion}

\begin{figure}
	\centering
	\includegraphics[width=0.9\columnwidth]{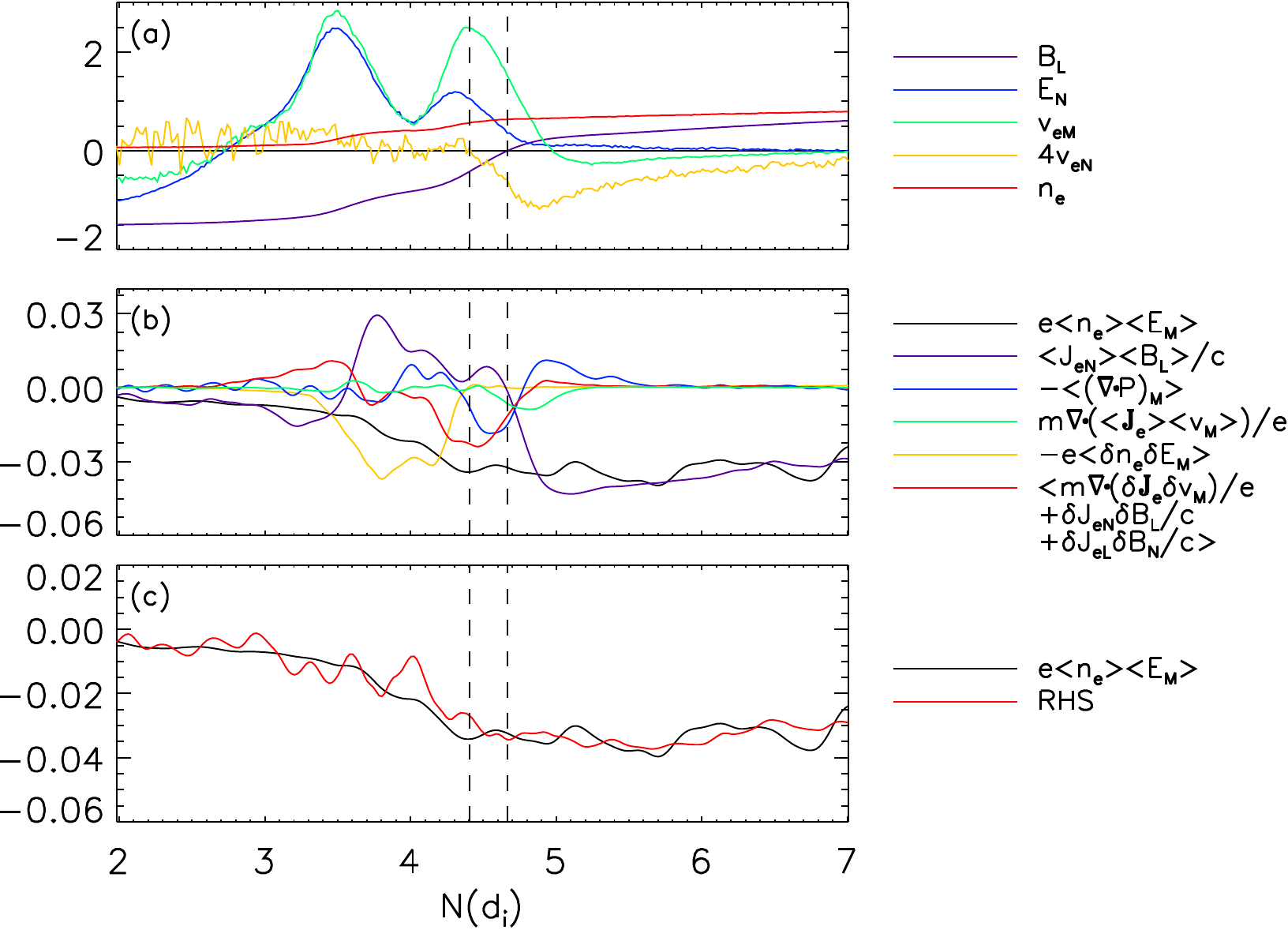}
	\caption{\label{ohmslate_100} Cuts along the N direction
          through the X line at $\Omega_{ci}t=38$. (a) $B_L$, $E_N$,
          $v_{eM}$, $v_{eN}$, and $n_e$ where the vertical dashed lines
          mark the X line and stagnation point $v_{eN}=0$. (b) The
          various contributions from the $M$ component of the averaged
          Ohm's law. (c) Sum of the various contributions to the
          average of the right-hand side of Ohm's law in Eq.~(\ref{ohms}) compared
          to $e\langle n_e\rangle\langle E_M\rangle$. }
\end{figure}

As discussed previously, we have demonstrated that the turbulence that
develops during 3-D simulations of the MMS 16 October 2015,
reconnection event is strong enough to control the characteristic
layer widths at the magnetopause. We now address whether the
turbulence is strong enough to impact the effective Ohm's law
controlling large-scale reconnection. Our previous analysis of
simulations of this event \citep{price16a} considered the effects of
the turbulence on reconnection by evaluating the contributions of
various terms to an averaged Ohm's law measured within the electron
diffusion region.  The $M$ component of Ohm's law (the electron
equation of motion) is as follows:
\begin{equation}\label{ohms}
E_M =
-\frac{1}{c}(\mathbf{v_e}\boldsymbol{\times}\mathbf{B})_M-\frac{1}{ne}
(\boldsymbol{\nabla\cdot}\mathbb{P}_e)_M-\frac{1}{e}m_e
\mathbf{v_e}\boldsymbol{\cdot\nabla}v_{eM}.
\end{equation}
Here $m, n, \mathbf{v_e}$ and $\mathbb{P}_e$ are the electron mass,
density, velocity, and pressure tensor. Because the temporal evolution
of the turbulence is over a short time scale compared with the time
associated with large-scale reconnection, large-scale reconnection is
controlled by the Ohm's law that is averaged over the turbulence. This
assumption is normally satisfied since the turbulence is at the
$\rho_e$ scale with a frequency $\Omega_{\text{lh}}$, while large-scale
reconnection takes place on time scales longer than the Alfv\'en
transit time across the computational domain. For 3-D simulations
the average is evaluated by averaging Ohm's law over the $M$
direction. As discussed earlier for the fluctuating $E_M$, we carry
out this average by separating each quantity in Ohm's law into fluctuating
and averaged quantities, $f=\langle f\rangle+\delta f$ and then
averaging Eq.~\ref{ohms} over $M$. In addition to contributions
independent of the fluctuations (the usual laminar contributions to
Ohm's law) are terms quadratic in the fluctuations that correspond to
the anomalous drag $-\langle\delta n_e\delta E_M\rangle$ and anomalous
viscosity $\langle m{\bf\nabla}\cdot (\delta{\bf J}_e\delta
v_M)/e+\delta J_{eN}\delta B_L/c+\delta J_{eL}\delta
B_N/c\rangle$. The conclusion from earlier simulations of this event
\citep{price16a,le17a} was that the anomalous terms were important
during the early phase of reconnection. However, \citet{le17a} found
that the turbulence weakened at late time so that the anomalous terms
were no longer important. In Fig.~\ref{ohmslate_100}a we show late
time ($\Omega_{ci}t=38$) cuts of various parameters from the
simulations ($B_L$, $E_N$, $v_{eM}$, $v_{eN}$, and $n_e$) versus $N$ in
a cut through the X line. The vertical dashed lines correspond to the
locations of the X line and the electron stagnation point $v_{eN}=0$,
which are displaced \citep{cassak07a}. In Fig.~\ref{ohmslate_100}b are
the various contributions from the averaged Ohm's law. The electron
diffusion region is the entire domain where the dominant laminar terms
$e\langle n_e\rangle\langle E_M\rangle$ and $\langle
J_{eN}\rangle\langle B_L\rangle/c$ do not balance. This region extends
well past the magnetosphere side of the stagnation point. The laminar
pressure tensor term which dominates the 2-D simulations continues to
be significant in the 3-D system. The anomalous viscosity term is large
across the entire region between the X line and stagnation point, while
the drag contribution is large on the magnetosphere side of the
stagnation point. Finally, in Fig.~\ref{ohmslate_100}c we show that
the total of all of the terms in the averaged Ohm's law balance the
average $E_M$ at this time. Thus, we reach a different conclusion than
\citet{le17a}. The turbulence remains strong enough to significantly
impact Ohm's law even at late time. The reason for the discrepancy is
unknown.

Before making further comparisons between the simulations and the MMS
data, we must establish the correspondence between the units used in
the simulation and those used in spacecraft measurements.  For the
asymptotic parameters of the 16 October 16 2015 event ($B_{L,\text{sh}}\sim23
\text{ nT}$, $B_{L,\text{ms}}\sim39 \text{ nT}$,
$n_{\text{sh}}\sim11.3/\text{cm}^3$, $n_{\text{ms}}\sim0.7/\text{cm}^3$) with ``sh''
and ``ms'' subscripts denoting the magnetosheath and magnetosphere
respectively, $d_{e,\text{sh}}\sim1.6\text{ km}$, $d_{i,\text{sh}}\sim68\text{ km}$,
$\omega_{ce,\text{sh}}\sim4.05\text{ kHz}$,
$\Omega_{\text{lh},\text{sh}}\sim95\text{ Hz}$, $\omega_{ci,\text{sh}}\sim2.2\text{
  Hz}$, $v_{A,\text{sh}}\sim150\text{ km/s}$ and $E_{0,\text{sh}}\sim3.4\text{
  mV/m}$.  In our simulations we find a reconnection electric field of
$\sim0.2\text{ mV/m}$ for either mass ratio, a value that would be
very difficult to detect observationally. In fact, MMS observations
reveal spikes in $E_M$ with much larger values, peaking around $\pm10
\text{ mV/m}$.  In addition, large amplitude, short-timescale
fluctuations of the parallel electric field $E_{\parallel}$, up to
$100\text{ mV/m}$, have been reported \citep{ergun16a, ergun17a}.
These intense parallel electric fields are not observed in our
simulations.

The question, then, is whether the MMS electric field measurements
correspond to an effective average over the turbulence in the
simulation or a slice at a particular value of $M$.  To answer this
question, note that the particle instruments on MMS directly measure
the full distribution function of electrons in $30\text{ ms}$ and of
ions in $150\text{ ms}$. The frequency $\omega$ of the fluctuations in
the simulation is around $\frac{1}{4}\Omega_{\text{lh},\text{sh}}=24\text{
  Hz}$ so the period of the waves is around $260\text{ ms}$. Thus, the
electron data are collected over a very short period compared to the
wave period. The MMS instruments are therefore measuring the local
electron Ohm's law and not the average Ohm's law that controls the
global reconnection rate.

To understand the challenge associated with deducing the global
reconnection rate directly from the MMS data, we translate the
reconnection rate determined from our simulation into real units. The
electric field in the simulation in SI units is normalized to
$E_0=B_{L,\text{sh}}c_{A,\text{sh}}$. For $B_{L,\text{sh}}=23 \text{ nT}$ and $n_{\text{sh}}=11.3/\text{cm}^3$,
$c_{A, \text{sh}}=150\text{ km/s}$ and $E_0=3.4\text{ mV/m}$. Thus, based on
Figure~\ref{ohmslate_100}c, we obtain the reconnection electric field
$E_{\text{rec}}=0.17\text{ mV/m}$. Extracting such a very small electric field
directly is problematic first because it is small and second because
the turbulent fluctuations in the electric field are typically of order
of $10\text{ mV/m}$ or higher
\citep{burch16a,ergun16a,ergun17a,graham17a}. Similarly, we can
translate the drag term in Fig.~\ref{ohmslate_100}b into the units
of an effective electric field, which yields $0.15\text{ mV/m}$. The
evaluation of the drag from direct measurements is a challenge because
it is necessary to carry out a time average of the correlation of the
product of the fluctuations. This was carried out earlier using THEMIS
magnetopause data \citep{mozer11a} and more recently using MMS data
\citep{graham17a}. The effective drag electric field was around
$0.5\text{ mV/m}$ in the more recent analysis from MMS. However, the average
was evaluated by simply averaging over the four spacecraft and using a
low-pass filter. Thus, the result was noisy and therefore probably not
very reliable. Further, the authors concluded that the drag terms were
small in comparison with the local values of the electric field and
therefore unimportant. As discussed previously, however, the drag
terms only apply to the analysis of the large-scale reconnection
electric field, which based on the simulation is of order
$0.17\text{ mV/m}$. Thus, on this basis the measured drag terms are large
enough to balance the large-scale reconnection electric field.

\begin{figure}
	\centering
	\includegraphics[width=0.9\columnwidth]{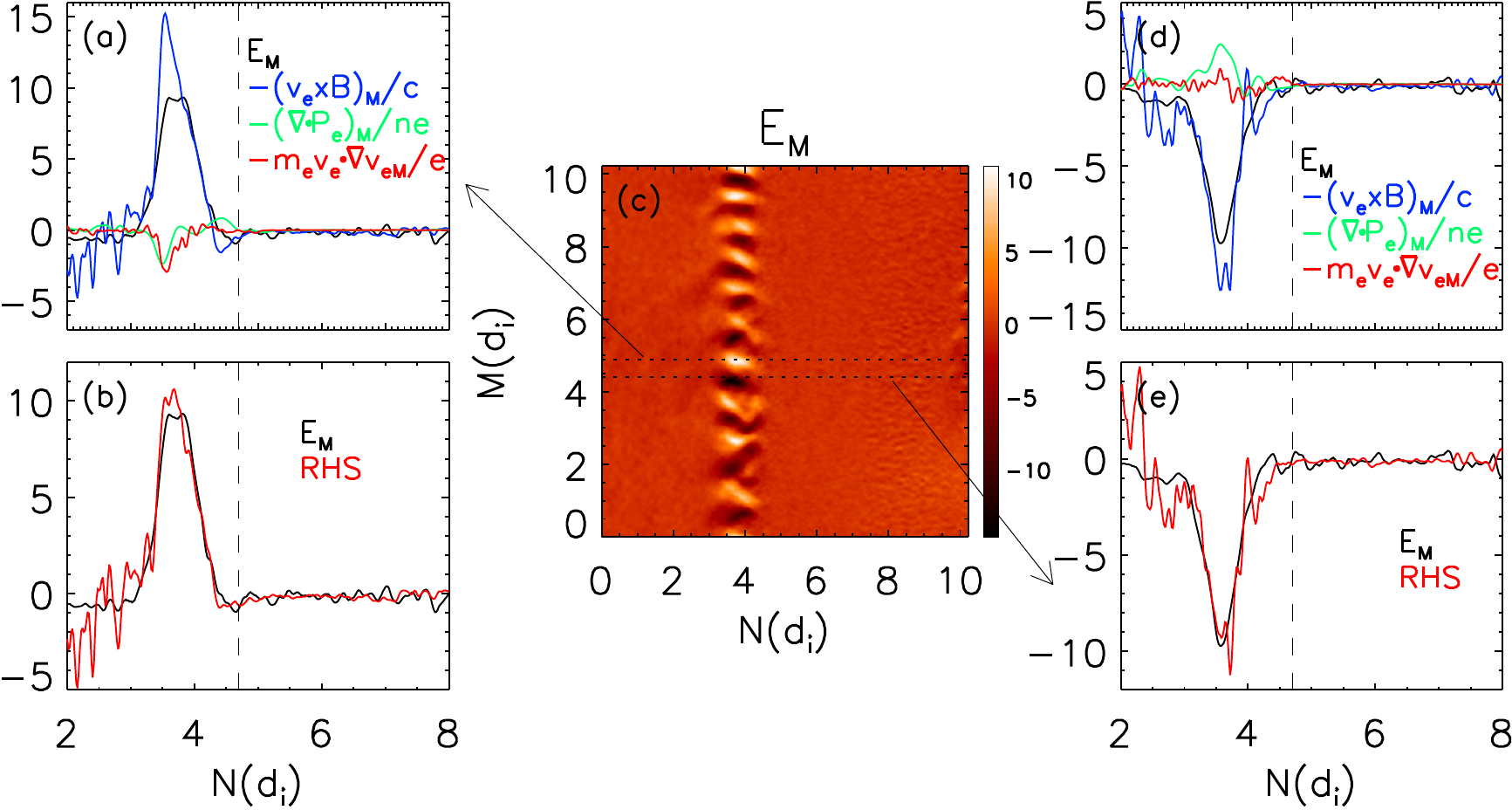}
	\caption{\label{ohmscut} Evaluation of Ohm's law on cuts
		through the region of instability.  For direct comparison to
		data, values are converted from our normalized units to
		mV/m.  (a) The terms in Ohm's law from equation
		\eqref{ohms} for a cut through $M=4.9$.  (b) The sum of
		the left and right sides of equation \eqref{ohms} for $M=4.9$.
		(c) $E_M$ in the $M-N$ plane at $t=38$.  The
		horizontal dotted lines denote locations of two cuts, at
		$M=4.4$ and $4.9$. (d) The terms in Ohm's law from
		equation \eqref{ohms} for a cut through $M=4.4$.  (e)
		The sum of the left and right sides of equation \eqref{ohms}
		for $M=4.4$.  The vertical dashed lines in Figures~\ref{ohmscut}a--\ref{ohmscut}b and 
		\ref{ohmscut}d--\ref{ohmscut}e indicate the position of the X line.}
\end{figure}

In order to determine the structure of the local Ohm's law and
therefore what MMS would measure within the diffusion region, we
examine the various terms in the $M$ component of Ohm's law in
equation~(\ref{ohms}) in a cut through the X line. We emphasize that this does
not represent the actual time dependence of the measurements from MMS
but is meant to emphasize the significant differences between the
averaged Ohm's law and that from a local measurement. In Figure
~\ref{ohmscut} we present data from two sample cuts through the
electron diffusion region along the $N$ direction. Figure~\ref{ohmscut}c
shows $E_M$ near the X line in the $M-N$ plane at $t=38$.
% along the dashed line through the x-line of Fig.~\ref{jez}(?)
Figure~\ref{ohmscut}a displays the separate terms in Ohm's law (equation~(\eqref{ohms})) at
$M=4.9$ along a cut in the $N$ direction (the upper dashed line in
Figure~\ref{ohmscut}c). Figure~\ref{ohmscut}b shows $E_M$ and the sum of the terms from the
right-hand side of equation~(\ref{ohms}). The two curves are in close
agreement, which confirms that the simulation data are consistent with
momentum conservation based on the electron equation of motion. Note
also that the vertical scale is expressed in mV/m so the curves
reflect the size of the terms that should be visible in the MMS data.
Figures~\ref{ohmscut}d and \ref{ohmscut}e show the same information for a cut through
$M=4.4$.  The value of $E_M$ peaks around $\pm10\text{ mV/m}$, very
close to the values reported in the MMS data \citep{burch16a}.  The
peak value of $E_M$ changes sign between the two cuts, which are
separated by a distance roughly comparable to the distance between the
MMS spacecraft.  Interestingly, a similar difference in polarity is
seen in the MMS data (see Figure 5 of \citet{burch16a}).  It should be
emphasized that the large value of $E_M$ shown in these cuts is a
result of the turbulence and does not reflect the rate of magnetic
reconnection.  The reconnection electric field, while present, is 2
orders of magnitude smaller and can only be extracted by the type of
averaging discussed above.

\begin{figure}
	\centering
	\includegraphics[width=0.9\columnwidth]{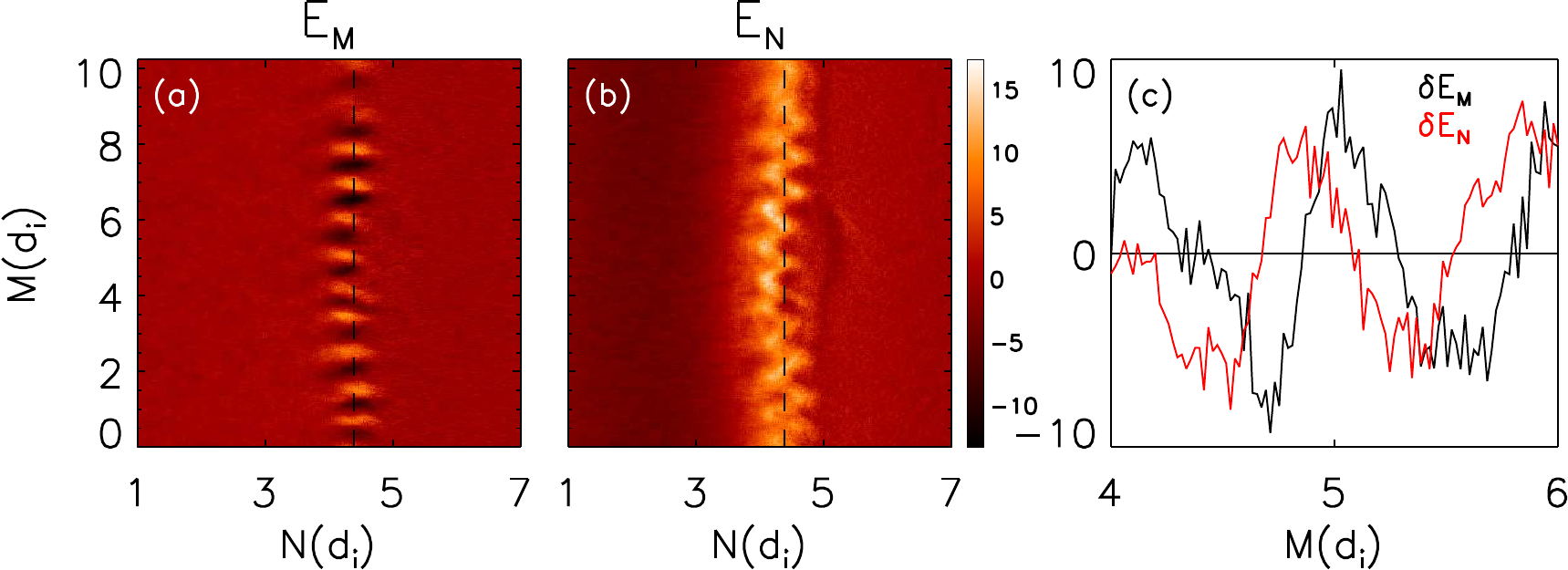}
	\caption{\label{eturb} (a) $E_M$ and (b) $E_N$ in the $M-N$ plane at the X line at $t=20$.
          For direct comparison to data, values are converted from our
          normalized units to mV/m.  (c) $\delta E_M$ and
          $\delta E_N$ through the vertical dashed line in Figures~\ref{eturb}a
          and \ref{eturb}b.  Figures~\ref{eturb}a and \ref{eturb}b are normalized to the same
          value.}
\end{figure}

As a further demonstration that $E_M$ is primarily associated with the
turbulence, Figure \ref{eturb} shows $E_M$ and $E_N$ (Figures~\ref{eturb}a and
\ref{eturb}b) in the $M-N$ plane near the X line at $t=20$.  In Figure~\ref{eturb}c we
plot cuts of $\delta E_M$ and $\delta E_N$ at the locations denoted by
the vertical dashed lines in Figures~\ref{eturb}a and \ref{eturb}b.  As the current layer
breaks up, it naturally produces large values of $E_M$ as the large
electron currents in the $M$ direction are diverted into the $N$
direction. These $N$-directed flows are driven by $E_M$.  The fact
that $\delta E_M$ and $\delta E_N$ are similar in magnitude and
roughly $90^{\circ}$ out of phase indicates that the fluctuations are
linked and not due to a steady-state reconnection process. Of course,
the turbulence itself might undergo reconnection on faster time scales
and produce electric fields larger than the nominal value of
0.1~mV/m. Such a possibility requires further analysis and comparison
with observations.

Multiple MMS observations of magnetopause electron diffusion regions
have found features similar to those in Figure \ref{ohmscut}
\citep{ergun17a}.  Since the observed turbulence did not satisfy the
properties of homogeneous LHDI it was suggested that some other
mechanism was responsible.  However, the findings presented here
suggest that the governing instability has all of the characteristics
of a longer wavelength version of LHDI.  The instability has a
dominant wavelength satisfying $k\rho_e \approx
(m_eT_e/m_iT_i)^{0.25}$, is observed in both electric and magnetic
field components and has a wave vector that is dominantly, but not
strictly, perpendicular to the local magnetic field.  The frequency of
the instability falls in the range of frequencies unstable to LHDI,
$\omega_{\text{ci}} \leq \omega\leq\Omega_{\text{lh}}$ and the growth
of the instability is closely correlated with the steepening and
relaxing of a density gradient (and therefore the ion pressure
gradient, which is the basic driver of drift instabilities at the
magnetopause).  Similar instabilities have been seen in other
three-dimensional reconnection simulations (albeit with different
initial conditions) and were also attributed to LHDI
\citep{daughton03a, pritchett12a,le17a}.

%{\bf What about more recent papers?}

%However, the instability seen in \citet{ergun17a} does not

%appear to satisfy $\mathbf{k}\boldsymbol{\cdot}\mathbf{B}=0$, nor does

%it have a strong pressure gradient to drive the instability.  It is

%possible that this instability is not the same as ours, as we see a

%strong gradient in the electron pressure despite density and

%temperature gradients opposing each other.

%As further defense, include derivation of frequency of cusp-like

%motion of electrons causing crescents (which I don't have because it

%was on a chalkboard in MD), showing the frequency scales like

%$\Omega_{\text{lh}}$ and not $\Omega_{ce}$.  It is very likely that electrons

%resonate with these waves.  The ions can also easily resonate with the

%wave, as it is of higher frequency that $\Omega_{ci}$.  A resonance

%with both species is necessary for the mode to drive transport.  Most

%of this is copy-pasted from Jim's email to Bob, should be worded

%better.

\begin{acknowledgments}
This work was supported by NASA grant NNX14AC78G.  We have benefited
greatly from conversations with members of the MMS team. 
The simulations were carried out at the National Energy Research Scientific Computing
Center.  The data used to perform the analysis and construct the
figures for this paper are available upon request.

\end{acknowledgments}

%\end{article}


%\bibliographystyle{agufull08}
%\bibliographystyle{apsrev4-1}
%\bibliography{paper}

\end{document}